# Effect of the reaction medium on hydrolysis of sodium borohydride

N. KARAKAYA AKBAŞ[*] and B. KUTLU[**]

[*]Gazi Üniversitesi, Fen Bilimleri Enstitüsü, Fizik A.B.D., Ankara, TURKEY

[**]Gazi Üniversitesi, Fen Fakültesi, Fizik A.B.D., Ankara, TURKEY

**Abstract**

In this study, he effect of the reaction medium on the initial hydrogen decomposition of the sodium borohydride (SBH) hydrolysis reaction in the presence of the Co catalyst surface has been theoretically investigated using the Castep package program bottomed on DFT. The aim of the research is to distinguish the catalytic effects of fcc cobalt surfaces on the decomposition of hydrogen from SBH and to identify the effects of the environments made by the hydrolysis of $H_2O$ at the most efficient the surface on H decomposition. Therefore, the common surface types of the face-centered cobalt crystal were used to determine the effect of fcc surface structure on H extrication from SBH and $H_2O$ extrication. The activation energies required for H extrication from SBH on the catalyst surface with the lowest activation barrier were determined for each possible reaction medium containing different density of chemical species. According to our results, the value of activation energy increases as the ratio of OH radicals increases.

**Introduction**

As a result of traditional methods used in energy consumption, environmental pollution has shown that it is very important to use clean energy sources instead of fossil fuels that cause global warming. One of the alternative sources of clean energy to carbon-based fuels is hydrogen. Although hydrogen has a higher energy density per unit mass than gasoline, the volumetric energy density is quite low. Because of this reason, various techniques to store hydrogen such as compressed-cryogenic tanks and sorbent materials that exploit physical interactions are far from fulfilling the requirements [1-3]. Chemical hydrides are promising materials that hold hydrogen at room temperature through chemical bonds [3]. Although water is an important source for hydrogen production, metal borohydrides ($MBH_4$) contains more hydrogen than water, so recent studies on hydrogen production have focused on the hydrolysis reaction of 1 mol of metal borohydride in 2 moles of water. This reaction is capable of producing 4 moles of $H_2$. Recently, experimental studies on the production of hydrogen from metal borohydrides (MBH4) have shown that the hydrolysis rate of catalytic MBH4 reactions depends on the proportion of OH radicals released from the decomposition of water [4,5]. So far, relatively few amounts of research have considered this effect clearly from the theoretical manner [6].

Thermolysis [8], hydrolysis [4], and its synchronous consideration in the alike reaction environment [9,10] are existing process for the hydrogen release stock in SBH. Among these techniques, the hydrolysis method has been reported to be more efficient [7,11]. The optimal hydrolysis of SBH was descriptioned [4] as;

$$NaBH_4 + 2 H_2O \rightarrow NaBO_2 + 4 H_2$$

Experimental results show that sodium metaborate is highly stable during the reaction and also increases the pH of the solution due to water decomposition. In recent years, experimental studies have shown that Co is one of the most active metal catalysts for the NaBH4 hydrolysis reaction of [4,12]. To date, studies on this reaction have shown that the reaction is exothermic and hence needs no energy input. The products (metaborates) of reaction are environmentally safe. The volumetric and gravimetric hydrogen storage efficiencies are high compared to other chemical hydrides and pure $H_2$ can be generated even at low temperatures [13]. The Co-based catalysts have good catalytic performance and there are considerable experimental studies have been conducted until today. Also, some researches emphasize the importance of cycle counts for catalyst and rate of degradation in practical uses of hydrogen storage.

Also, water used in the production of hydrogen is a requirement in terms of the solubility of reactants and by-products. Understanding the reaction kinetics by reducing the amount of water on the pathway to obtaining $NaBO_2$ during the reaction of SBH with water is important to achieve the high energy density of the hydrogen storage system. The coherence between the theoretical and experimental studies guides development of new hydrogen storage materials and research of their catalytic activity. It should be considered that the differences of experimental investigations, such as molarities of reactants, temperature, etc., prevent an actual comparison with theoretical studies however they still give hints for the new researches.

The aim of this study is to determine the preferable cobalt surface for the inseption way of NaBH4 hydrolysis and decomposition of water. For this purpose; separation of water, hydrogen decomposition from NaBH4 and changes in environment terms due to this were investigated.

In summary, we examined the mechanism of environment differences that may occur during the SBH hydrolysis reaction on low index cobalt surfaces.

$$H_2O \rightarrow H + OH \tag{2}$$

$$NaBH_4 \rightarrow NaBH_3 + H \tag{3}$$

$$NaBH_4 + m\, H_2O \rightarrow NaBH_3 + H + m\, H_2O \tag{4}$$

$$NaBH_4 + (m-1)[H_2O + H + OH] \rightarrow NaBH_3 + H + (m-1)[H_2O + H + OH] \tag{5}$$

$$NaBH_4 + m\,(H + OH) \rightarrow NaBH_3 + H + m(H + OH) \tag{6}$$

where m=2. The second aim of this study is to investigate the effects of species to the SBH hydrolysis reaction. For this purpose, activation energies of the inception way, which may have different possibilities, of the catalytic SBH hydrolysis reaction on the surface were determined using DFT calculations in the presence of species. It is possible to define the inception way in the SBH hydrolysis reaction on the surface by the following reaction equations to create different environment terms.

These proposed reactions correspond to different reaction combinations for the inception way, where one of the hydrogens in SBH decomposes. The last two are related to the decomposition of water. In this study, the effect of possible environmental terms on the decomposition of one of the hydrogen in SBH during the hydrolysis reaction of SBH was theoretically investigated.

**Computational Method**

In this study, calculations were used Castep programme[14,15]. GGA/PBE functional was preferred as exchange correlation potential[16,17].

The low index Co surfaces were modeled using a four-layer slab and 3x3 unit cell. Of the 4 layers, the lower two layers are fixed in such a way that they retain their position in the crystalline structure, while the upper two layers is released. After that optimization calculates was made for reactions, transition state calculations were carried of the reactions.

All calculations were performed using cutoff energy of 500 eV and the k-point was set to be 4x4x1. Convergence parameters of 5.0×10-5 eV/atom for energy, 0.1 eV/Å for force, 0.005 Å for displacement and 0.2 GPa for stress were used. The vacuum width was chosen as 20 Å and taking into account the spin-polarizing effect in calculations.

The reaction energy (ΔH),

$$\Delta H = E_{Pr./surf.} - E_{Rc./surf.} \quad (7)$$

can be given as here, $E_{Rc/surf.}$ of reactants and $E_{Pr/surf.}$ of the products are the total energy where ΔH is the enthalpy of the reaction. A negative enthalpy value indicates that a reaction is exothermic. Otherwise it is called an endothermic reaction.

The forward activation energies ($E_{FA}$) were defined using the total energy of molecules on the surface. The values of $E_{FA}$ is calculated by

$$E_{FA} = E_T - E_R \quad (8)$$

where $E_T$ is transition state energies of system, $E_R$ is energy of system formed from the reactants and the surface and $E_P$ is total energy of products and the surface.

Activation energies of decomposition of chemical species on various metal surface were successfully obtained by performing transition states calculations [18]. In this study, we made TS calculations for each reaction step and for this, we were used the quadratic synchronous transit (**QST**) process in the Castep programme [19].

**Mechanisms of Hydrolysis of NaBH$_4$**

*Energy barrier calculations for fcc cobalt surfaces*

First of all, Co(111), Co(100), and Co(110) surfaces of metallic fcc cobalt, which is used as catalyst, were modeled by four layers of metals and a p(3x3) supercell. Initial and final states of possible decomposition reactions based on the main reaction are modeled and optimized separately for the surfaces. LST/QST transition state calculations were made using the initial and final states.

Table I. Forward energy barriers, reverse energy barriers and enthalpy of reactions given by Equations 2-4. Calculations were performed on low index cobalt surfaces.

| | Basic Paths | $E_{FA}(eV)$ | $E_{RA}(eV)$ | $\Delta H(eV)$ |
|---|---|---|---|---|
| 111 | $H_2O \to H + \cdot OH$ | 1,10 | 1,49 | -0,39 |
| | $NaBH_4 \to NaBH_3 + H$ | 0,26 | 0,63 | -0,37 |
| | $NaBH_4 + m\ H_2O \to NaBH_3 + H + m\ H_2O$ | 0,21 | 1,06 | -0,85 |
| 100 | $H_2O \to H + \cdot OH$ | 1,14 | 1,85 | -0,71 |
| | $NaBH_4 \to NaBH_3 + H$ | 0,41 | 1,02 | -0,61 |
| | $NaBH_4 + m\ H_2O \to NaBH_3 + H + m\ H_2O$ | 0,38 | 1,59 | -1,21 |
| 110 | $H_2O \to H + \cdot OH$ | 1,08 | 1,82 | -0,74 |
| | $NaBH_4 \to NaBH_3 + H$ | 0,41 | 0,98 | -0,57 |
| | $NaBH_4 + m\ H_2O \to NaBH_3 + H + m\ H_2O$ | 0,37 | 1,70 | -1,33 |

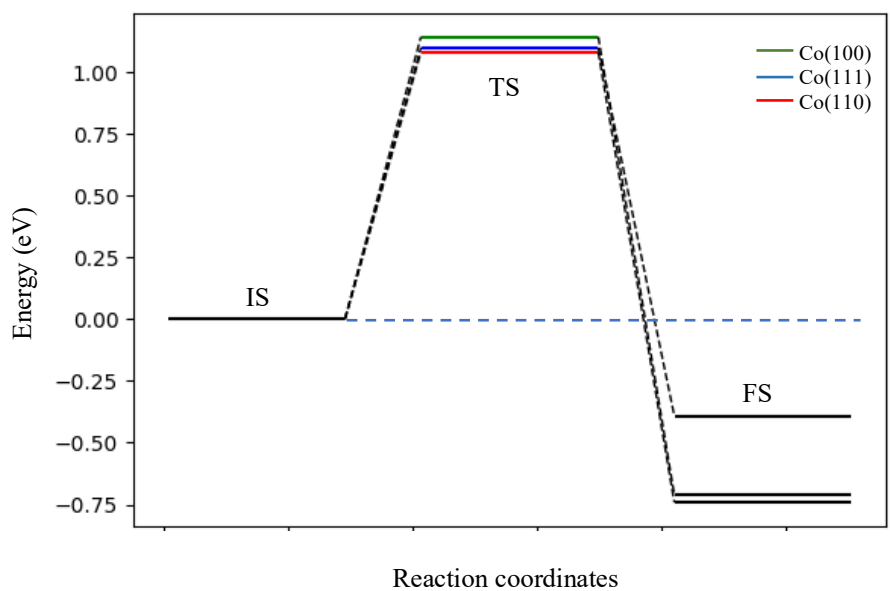

(a)

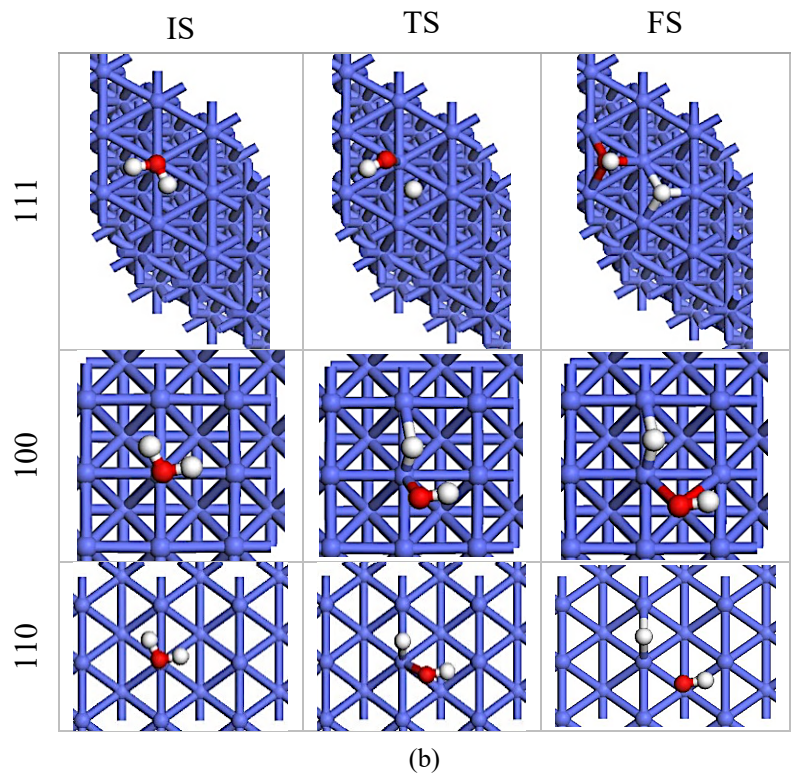

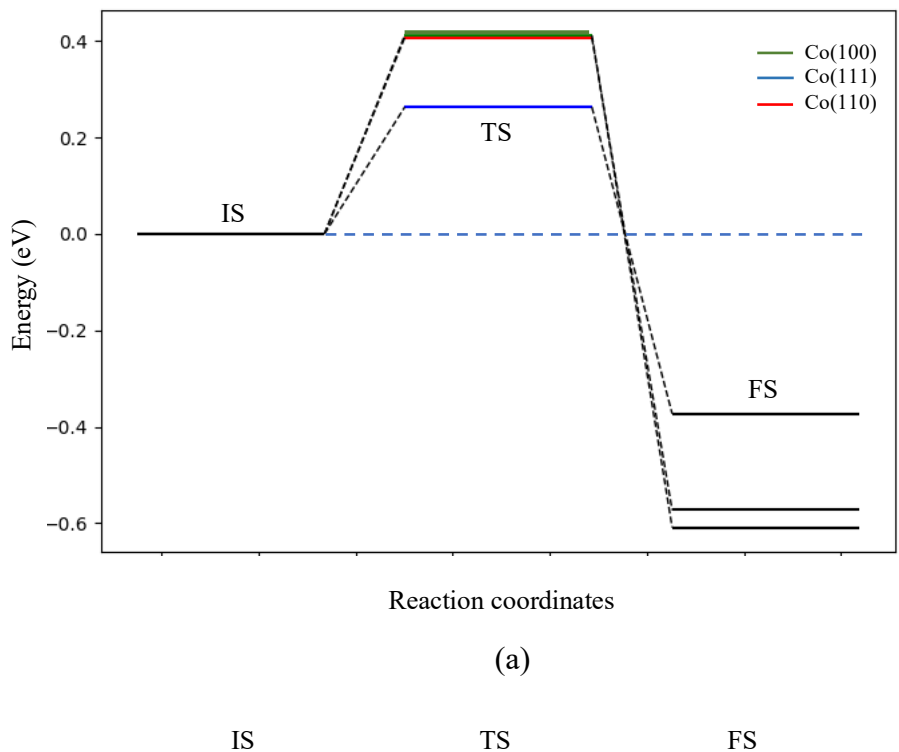

Fig.1 (a) Energy profile and (b) configurations of the initial state, transition state, and final state on the cobalt surfaces for the $H_2O \rightarrow H + \cdot OH$ reaction.

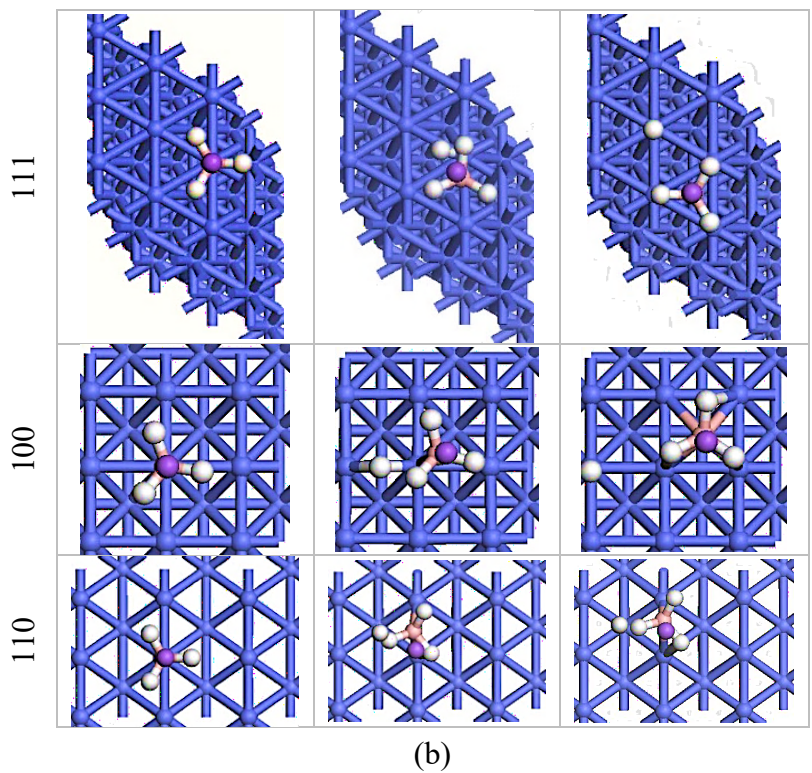

(b)

Fig.2 (a) Energy profile and (b) configurations of the initial state, transition state, and final state on the cobalt surfaces for the NaBH$_4$ → NaBH$_3$ + H reaction.

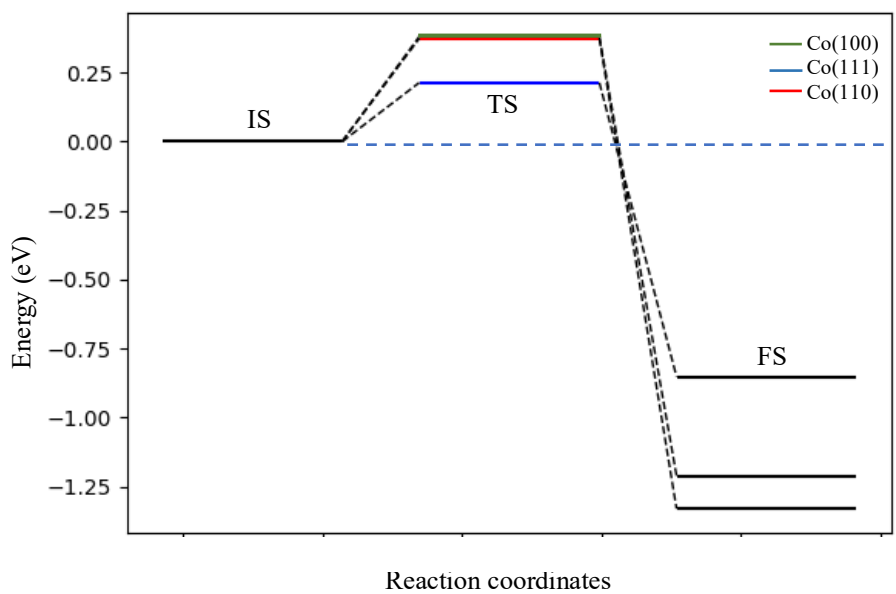

(a)

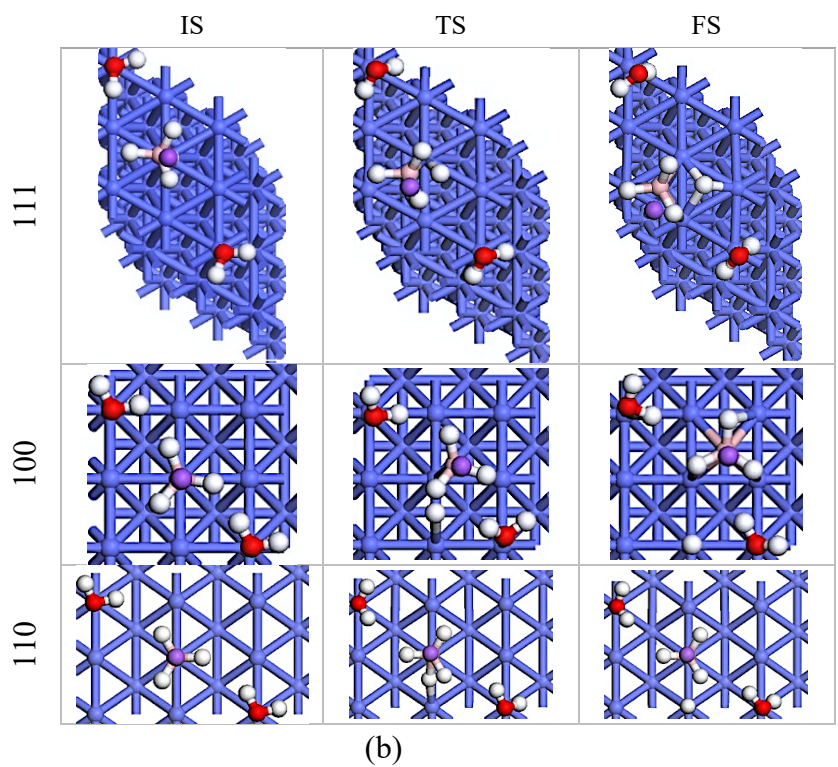

(b)

Fig.3 (a) Energy profile and (b) configurations of the initial state, transition state, and final state on the cobalt surfaces for the $NaBH_4 + m\ H_2O \rightarrow NaBH_3 + H + m\ H_2O$ reaction.

LST/QST calculations do not differ in activation barrier values required for molecular water decomposition. And also; The values of energy required for hydrogen decomposition from $NaBH_4$ are shown in Table I for the anhydrous medium. As can be seen from Table I, the lowest activation energy for the decomposition of one of the hydrogen in $NaBH_4$ in the aqueous and non-aqueous environment was obtained on the Co(111) surface. These results showed that the Co(111) surface from fcc Co surfaces is a preferred surface for catalytic effect in hydrogen separation from $NaBH_4$. This result is in agreement with a study by Rostamikia et al.[20]. They calculated the activation barrier required for the decomposition of $BH_4 \rightarrow BH_3 + H$ on the Au(111) surface as 0.37 eV[20]. However, hydrolysis of water takes place at fcc Co surfaces with energies very close to each other. Therefore, it can be said that there is no preferred fcc Co surface for the hydrolysis of water.

*Transition state search calculations for Co(111)-(4x4) surface*

The calculations in this section are designed to determine the transition states of the reactions given in Equations (2-6). The Co(111) surface is created using a four-layer (4x4) surface where the top two layers are allowed to relax and the bottom two layers are constrained by the stack positions.

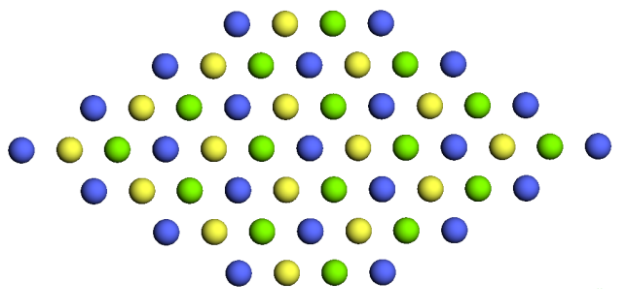

Fig.4 The possible adsorption sites of fcc cobalt surfaces. The positions of the top, fcc and hcp are the blue, green, and yellow balls, respectively. The bridge position is the midpoint between the two blue balls.

There are four different adsorptive sites for Co(111)-(4x4) surface morphology; top, hcp, fcc, and bridge (shown in Fig.4). Before the transition state calculation, the reactants and products given in the reaction steps were optimized on the 4x4 surface. When the reactants and products of the reaction are optimized; $NaBH_4$, $H_2O$ are adsorbed on top sites, $NaBH_3$ is adsorbed on the hcp site. In presence of other adsorbents, H and ·OH adsorption sites are very high and weakly bound although H and ·OH are adsorbed on top site of the clean surface.

To determine the effects of $H_2O$, ·OH and H species for the decomposition of $NaBH_4$ on the surface, we first obtained transition states for all reaction steps.

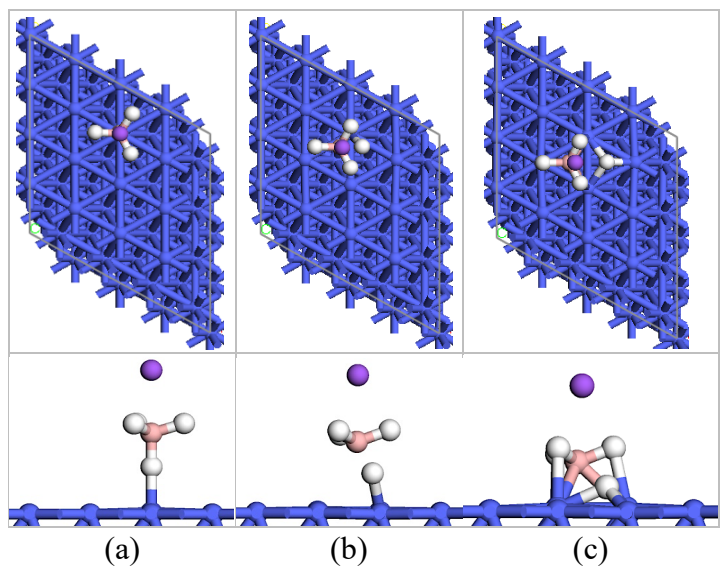

(a) (b) (c)

Fig.5 The steady state configuration of (a) initial state (b) transition state and (c) final state on the Co(111) surface of Equation 3.

To determine the effects of ·OH species made by the decomposition of $H_2O$ molecules, the activation energy was obtained for the reaction given by the Equation (3) on the clean cobalt surface. The species which both reactants and products on the surface were optimized.

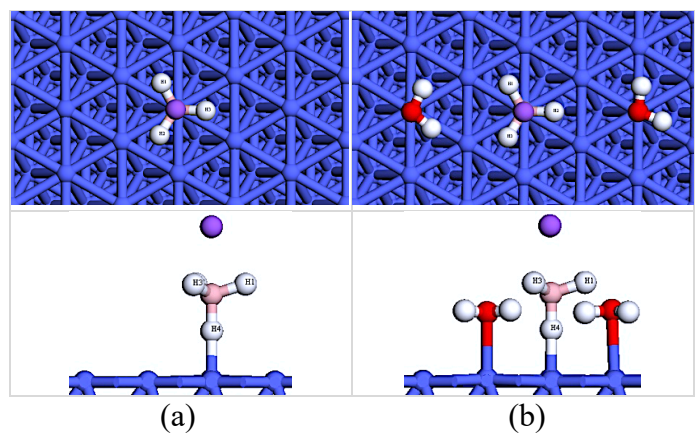

(a)                                    (b)

Fig.6 The steady state configuration of $NaBH_4$ molecule on the Co(111) surface in non-aqueous medium (a) and the presence of 2-mole water (b). (O, B, Na, H atoms are red, pink, purple, white respectively.)

As a result of the optimization, it was observed that the $NaBH_4$ molecule has minimum energy at the top position while the positions of $NaBH_3$ and H species after decomposition were configured to bound at the nearest fcc position and hcp position, respectively. $NaBH_4$ was optimized both in the presence and absence of water. It was seen that BH bond lengths in the $NaBH_4$ molecule have changed because of the interaction with the water. While the values of B-H1, B-H2, B-H3 bond lengths in the aqueous and anhydrous medium are 1.23 Å, the value of B-H4 bond length increases from 1.25 Å in the absence of water to 1.27 Å in the presence of water. This result indicates that water facilitates the separation of H from the $NaBH_4$ molecule.

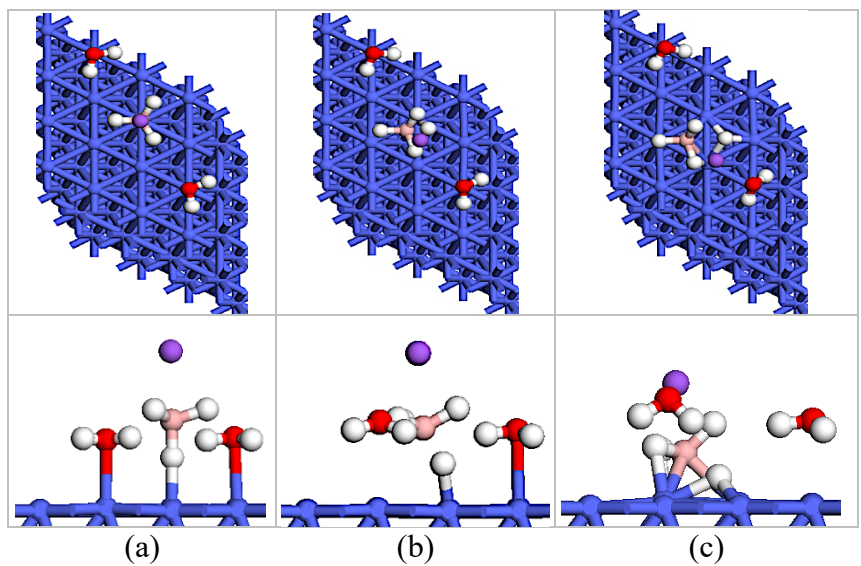

Fig.7 The steady state configuration of (a) initial state (b) transition state and (c) final state on the Co(111) surface of the PW3 possible reaction pathway.

On the other hand, the optimization result for the reaction in Equation (4) which can be considered as the starting point of the reaction for 1 mol $NaBH_4$ molecule seperate in 2 mol $H_2O$ in the catalytic environment is given in Fig.7 (a), (b) and (c). Additionally, we investigated the hydrolysis reaction of sodium borohydride in the presence of H and ·OH species. The variation in numbers of H and ·OH species changes the energetic behavior of the relevant reaction step. This reaction design which can be seen in Equations 4-6 represents different reaction environment. It is very well known that half of the hydrogen atoms come from water decomposition in the hydrolysis reaction. It means that water molecules have to decompose and release H and ·OH species to the reaction environment.

As seen in Table II, the reaction barrier energies for dissociation of $H_2O$ molecule were calculated as 1,121 eV in the PW1 step for a clean surface, 0.98 eV, and 1.13 eV in the PW8 and PW9 steps for the reaction environment, respectively. Our calculations are consistent with the reaction barrier energies obtained for the separation of water on various clean surfaces presented in many studies [21, 22-28]. The calculations of $H_2O$ decomposition on different Co surfaces by F. F. Ma et al. [22] showed that the reaction barrier for the O-covered Co surface is lower than the clean surface. Our results for the clean and non-clean Co surface supports the results of F.F. Ma et. al. [22].

Table II. Forward energy barriers, product energy barriers, and enthalpies of possible pathways for the first hydrogen dissociation from $NaBH_4$ in the catalytic environment. (m=2)

| | Pathways | | $E_{RB}$ | $E_{PB}$ | $\Delta H$ |
|---|---|---|---|---|---|
| **PW1** | $H_2O \rightarrow H + OH$ | TS1 | 1.12 | 1.51 | -0.39 |
| **PW2** | $NaBH_4 \rightarrow NaBH_3 + H$ | TS2 | 0.39 | 1.13 | -0.73 |
| **PW3** | $NaBH_4 + m\ H_2O \rightarrow NaBH_3 + H + m\ H_2O$ | TS3 | 0.34 | 1.99 | -1.66 |
| **PW4** | $NaBH_4 + (m-1)\ [H_2O+H+OH] \rightarrow NaBH_3 + m\ H + (m-1)\ [H_2O+OH]$ | TS4 | 0.35 | 2.01 | -1.67 |
| **PW5** | $NaBH_4 + (m-1)\ [H_2O+OH] \rightarrow NaBH_3 + (m-1)\ [H+OH+H_2O]$ | TS5 | 0.37 | 2.19 | -1.81 |
| **PW6** | $NaBH_4 + m\ [H+OH] \rightarrow NaBH_3 + (m+1)\ H + m\ OH$ | TS6 | 0.61 | 1.33 | -0.72 |
| **PW7** | $NaBH_4 + m\ OH \rightarrow NaBH_3 + (m-1)\ H + m\ OH$ | TS7 | 0.69 | 1.14 | -0.45 |
| **PW8** | $NaBH_4 + m\ H_2O \rightarrow NaBH_4 + (m-1)\ [H+OH+H_2O]$ | TS8 | 0.98 | 0.40 | -0.58 |
| **PW9** | $NaBH_4 + (m-1)\ [H+OH+H_2O] \rightarrow NaBH_4 + m\ [H+OH]$ | TS9 | 1.13 | 0.57 | -0.56 |

On the other hand, the transition state energies of the different atomic and molecular environment to the hydrogen abstraction from $NaBH_4$ molecule are given in Table II and Fig.8., respectively. The PW3 step given in Table II is the most probable reaction at the beginning of the reaction. The other reaction steps show possible steps that may occur after the beginning of the reaction. In this study, TS values for $H_2O$ decomposition are in good agreement with the results of other studies, despite the different adsorbents on the surface. The cause of the TS9 being higher than the TS8 is due to the presence of excess OH radicals in the medium in the PW9 reaction.

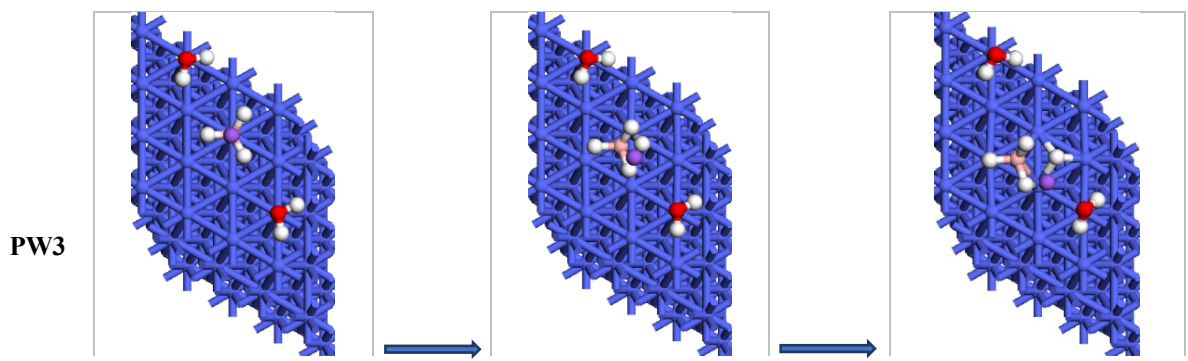

PW3

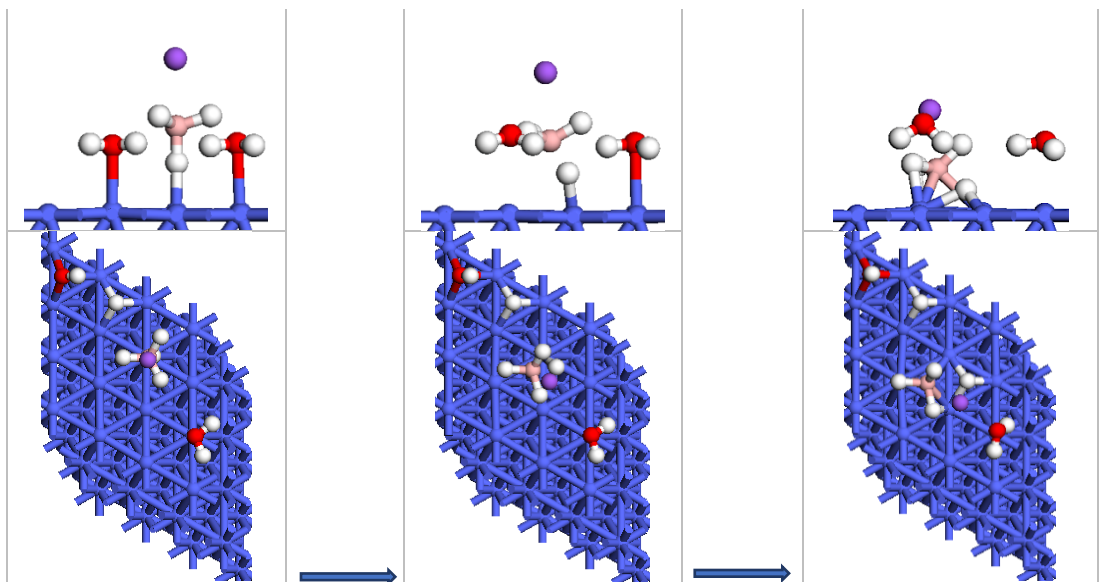
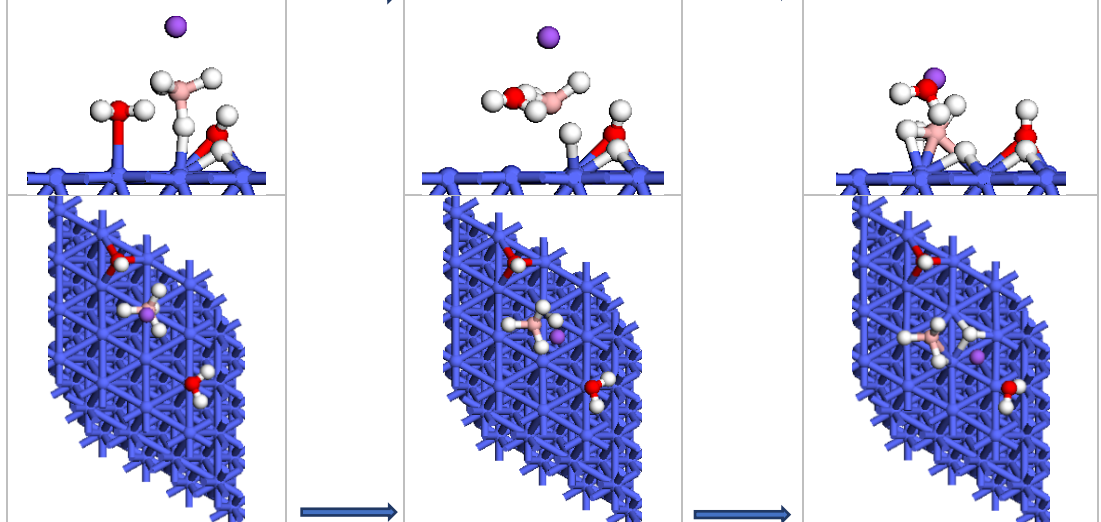
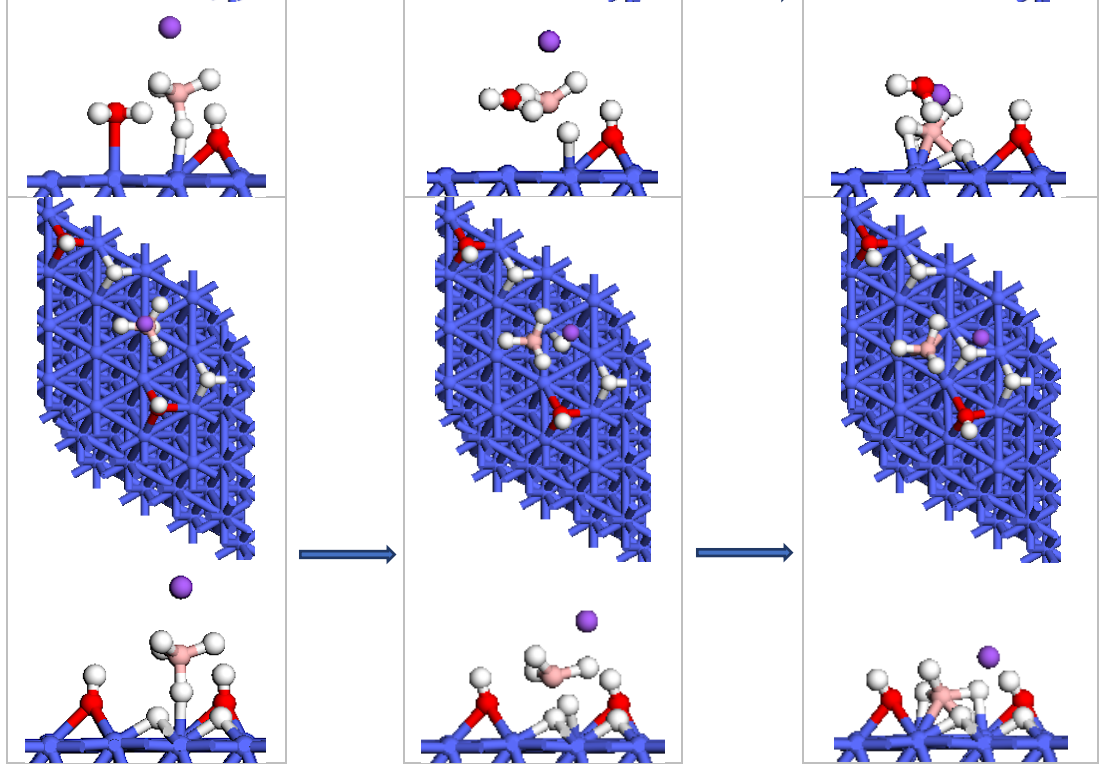

**PW4**

**PW5**

**PW6**

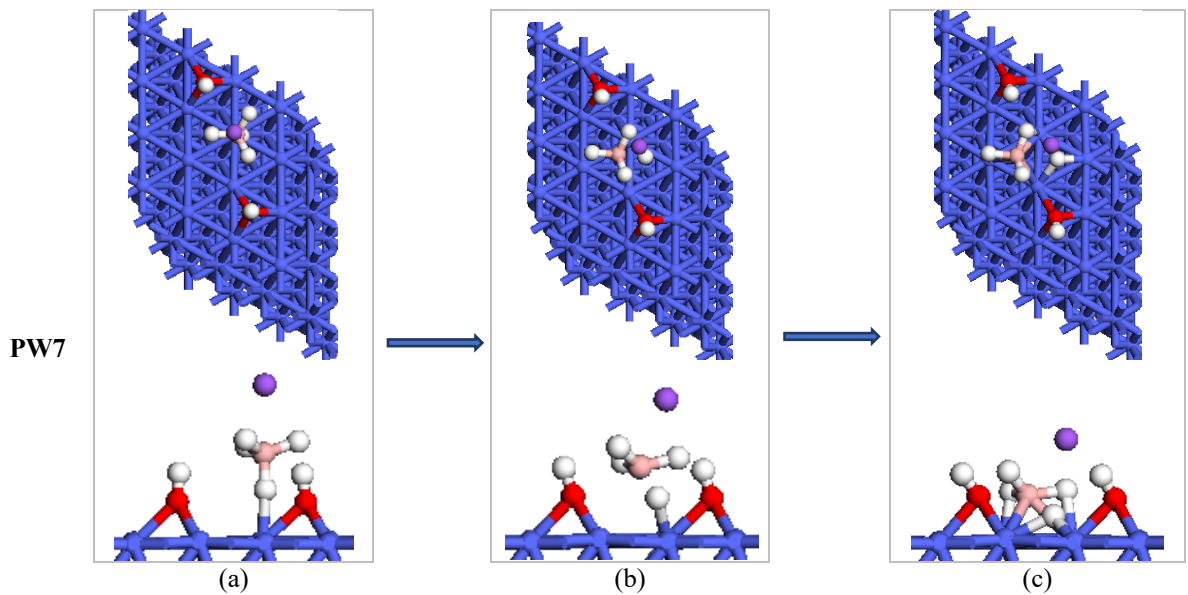

Fig.8. The steady state configuration of (a) initial state (b) transition state and (c) final state on the Co(111) surface of the PW3-PW7 possible reaction pathways.

The initial step of the reaction is given in Equation (1) was examined by the possible reaction paths PW3-PW7 for the dehydration of $NaBH_4$ for different OH radical concentrations. In the PW3 reaction pathway where there are no OH radical, the energy barrier is 0.337 eV and reaction energy is -1.66 eV, while in the PW4 reaction where 1 mol $H_2O$ is dissociated, the barrier is 0.354 eV and -1.67 eV. In the PW5 reaction pathway where an H atom is removed from the surface in the PW4 reaction pathway, the energy barrier increases to 0.374 eV. The energy barrier increased to 0.610 eV in the PW6 reaction path, where both $H_2O$ molecules were dissociated into ·OH and H. The activation energy barrier increases with increasing ·OH concentrations.

**Conclusions**

In this study, activation energies were obtained for possible reaction pathways for the separation of H from $NaBH_4$ molecule which is the first step in 1 mol $NaBH_4$ and 2 mol $H_2O$ reaction using DFT calculations. The variation of activation energy ($E_A$) corresponding to these steps is given in Fig.9.

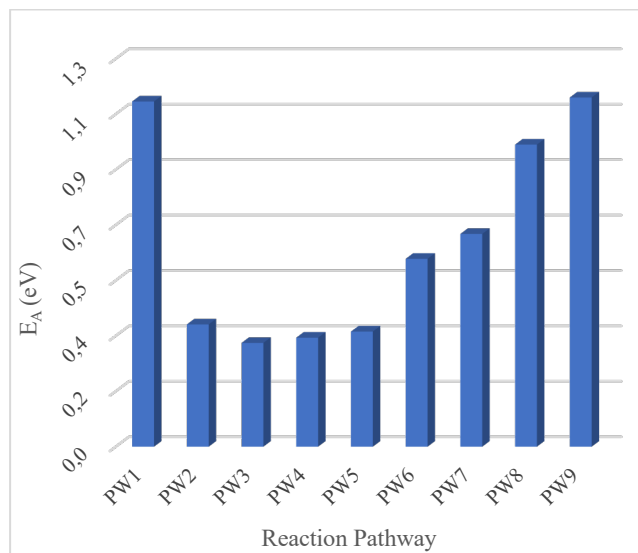

Fig.9. The variation of activation energy corresponding to reaction pathways.

As can be seen in Fig.9, the activation energy is increased for further reaction steps. The increase in activation energy is due to the formation of the ·OH in the reaction environment as a result of the decomposition of 2 mol $H_2O$. Acidic solutions have relatively more hydrogen ions, while alkaline (also called basic) solutions have relatively more hydroxyl ions. The acidity or alkalinity is determined by the amount of hydrogen ion (H+) or hydroxyl ($OH^-$) in the medium. pH is the measurement of acidity or alkalinity of the aqueous solution [29]. The results obtained for the energy barrier indicate that the reaction rate decreases as the basicity of the medium increases. This is in good agreement with experimental studies conducted by S.S. Muir et al [6]. The activation energies calculated for the PW5 and PW7 reaction pathways are greater than those obtained for the PW4 and PW6 reaction pathways where H+ are present. This indicates that the acidity of the environment decreases as H+ move away from the surface. For example, the activation energy for the PW4 is 0.354 eV, while for the PW5 it is 0.374 eV.

In this context, the change in the activation energies at the possible reaction stages are compatible with the concentration of the $OH^-$ ions inversely proportional to the reaction rate. In other words, the change in the activation energy is thought to result from the change in the pH of the medium. Also, the activation energy for the clean surface is greater than that obtained for the surfaces containing $H_2O$, whereas it is smaller than that of the surfaces where water completely decomposes to ·OH and H. This result indicates that the probability of formation of the first reaction step varies with the absorption of different molecules onto the catalytic surface. Moreover, the catalytic activity for the first reaction step is reduced by the absorption of different molecules onto the catalytic surface. Thus, it is possible to say that the OH radicals are effective in determining the rate of the first step corresponding to the dissociation reaction of H atom from $NaBH_4$. The DFT calculation results are in good agreement with the experimental results. The activation energy values obtained as a result of the transition state search calculations in the catalytic medium showed that the DFT calculations were distinctive in determining the effect of the intermediate states on the reaction rate.

## Acknowledgment

HPCC system by Gazi University was used for the calculative processes in this paper.

# References

1. M.P. Ramage, Transitions to Alternative Transportation Technologies; A Focus On Hydrogen, Washington, D.C., National Research Council; 2008. The National Academies Press. doi: 10.17226/12222.

2. U. Eberle, M. Felderhoff, F. Schuth, "Chemical and physical solutions for hydrogen storage", Angew Chem. Int. Ed., 48(2009):6608-6630.

3. Fuel cell technologies office multi-year research, development and demonstration plan, in: Hydrogen storage, Washington, D.C., U.S. Department of Energy; 2012.

4. H.I. Schlesinger, H.C. Brown, A.E. Finholt, J.R. Gilbreath, H.R. Hoekstra, E.K. Hyde, "Sodium borohydride, its hydrolysis and its use as a reducing agent and in the generation of hydrogen", J. Am. Chem. Soc., 75(1953):215-219.

5. A.M.F.R. Pinto, D.S. Falcão, R.A. Silva, C.M. Rangel, "Hydrogen generation and storage from hydrolysis of sodium borohydride in batch reactors", Int. J. Hydrogen Energy, 31(2006):1341-1347.

6. S.S. Muir, X. Yao, "Progress in Sodium Borohydride as a Hydrogen Storage Material: Development of Hydrolysis catalysts and reaction systems", Int. Journal of Hydrogen Energy, 36(2011): 5983-5997.

7. E.Y. Marrero-Alfonso, A.M. Beaird, T.A. Davis, M.A. Matthews., "Hydrogen generation from chemical hydrides", Ind. Eng. Chem. Res., 48(2009):3703–3712.

8. E. Fakioğlu, Y. Yurum, N.T. Veziroğlu, "A review of hydrogen storage systems based on boron and its compounds", Int. J. Hydrogen Energy, 29(2004):1371-1376.

9. E. Shaffirovich, V. Diakov, A. Varma, "Combustion-assisted hydrolysis of sodium borohydride for hydrogen generation", Int. J. Hydrogen Energy, 32(2007):207-211.

10. V. Diakov, M. Diwan, E. Shaffirovich, A. Varma, "Mechanistic studies of combustion-stimulated hydrolysis of sodium borohydride for hydrogen generation", Chem. Eng. Sci. 62(2007):5586-5591.

11. E.Y. Marrero-Alfonso, J.R. Gray, T.A. Davis, M.A. Matthews, "Minimizing water utilization in the hydrolysis of sodium borohydride: the role of sodium metaborate hydrates", Int. J. Hydrogen Energy, 32(2007):4723-4730.

12. P.K. Singh, T. Das, "Generation of hydrogen from NaBH$_4$ solution using metal-boride (CoB, FeB, NiB) catalysts", Int. J. Hydrogen Energy, 42(2017):29360-29369.

13. S.C. Amendola, S.L. Sharp-Goldman, M.S. Janjua, N.C. Spencer, M.T. Kelly, P.J. Petillo, "A safe, portable, hydrogen gas generator using aqueous borohydride solution and Ru Catalyst", Int. J. Hydrogen Energy, 25(2000):969-675.

14. M.C. Payne, M.P. Teter, D.C. Allan, T.A. Arias, J.D. Joannopoulos, "Iterative minimization techniques for ab initio total-energy calculations: molecular dynamics and conjugate gradients", Rev. Mod. Phys., 64(1992):1045-1097.

15. V. Milman, B. Winkler, J.A. White, C.J. Pickard, M.C. Payne, E.V. Akhmatskaya, R.H. Nobes, "Electronic Structure, Properties, and Phase Stability of Inorganic Crystals: A Pseudopotential Plane-Wave Study", Int. J. Quantum Chem., 77(2000): 895-910.

16. J.P. Perdew, K. Burke, M. Ernzerhof, "Generalized gradient approximation made simple", J. Physical Review Letters, 77(1996): 3865−3868.

17. M.D. Segall, P.J.D. Lindan, M.J. Probert, C.J. Pickard, P.J. Hasnip, S.J. Clark, M.C. Payne, "First-principles simulation: ideas, illustrations, and the CASTEP code", Journal of Physics: Condensed Matter, 14(2002): 2717-2744.

18. A. Akça, A.E. Genç, B. Kutlu, "BH$_4$ dissociation on various metal (111) surfaces A DFT study", Appl. Surf. Sci., 473(2019):681–692.

19. T.A. Halgren, W.N. Lipscomb, "The synchronous-transit method for determining reaction pathways and locating molecular transition states", Chem. Phys. Lett., 49(1977): 225–232.

20. M. Pozzo, G. Carlini, R. Rosei, D. Alfè, "Comparative study of water dissociation on Rh(111) and Ni(111) studied with first-principles calculations", J. Chem. Phys., 126(2007): 164706(1-12).

21. G. Rostamikia, A.J. Mendoza, M.A. Hickner, M.J. Janik, Journal of Power Sources, 196 (2011):9228-9237.

22. F.F. Ma, S.H. Ma, Z.Y. Jiao, X.Q. Dai, "Adsorption and decomposition of $H_2O$ on cobalt surfaces: A DFT study", Appl. Surf. Sci., 384(2016):10–17.

23. Z. Jiang, L. Li, M. Li, R. Li, T. Fang, "Density functional theory study on the adsorption and decomposition of $H_2O$ on clean and oxygen-modified Pd(100) surface", App. Surf. Sci., 301(2014): 468-474.

24. R.R.Q. Freitas, R. Rivelino, F.B. Mota, C.M.C. Castilho, "Dissociative adsorption and aggregation of water on the Fe(100) surface: A DFT study", J. Phys. Chem. C, 116(2012):20306–20314.

25. Z. Jiang, M. Li, T. Yan, T. Fang, "Decomposition of $H_2O$ on clean and oxygen-covered Au(100) surface: A DFT study", Appl. Surf. Sci., 315(2014): 16–21.

26. Y.Q. Wang, L.F. Yan, G.C. Wang, "Oxygen-assisted water partial dissociation on copper: a model study", Phys. Chem. Chem. Phys., 17(2015): 8231–8238.

27. A. Mohsenzadeh, K. Bolton, T. Richards, "DFT study of the adsorption and dissociation of water on Ni(111), Ni(110) and Ni(100) surfaces", Surf. Sci., 627(2014):1–10.

28. J.L.C. Fajín, M.N.D.S. Cordeiro, J.R.B. Gomes, "Density functional theory study of the water dissociation on platinum surfaces: General trends", J. Phys. Chem. A, 118(2014): 5832–5840.

29. R.D. Down, J.H. Lehr, "Environmental Instrumentation and Analysis Handbook, in: J.R. Gray, pH Analyzers and Their Application", 2005, p:460.